# Text Classification Using Data Mining

S. M. Kamruzzaman[1]
Farhana Haider[2]
Ahmed Ryadh Hasan[3]

## Abstract

Text classification is the process of classifying documents into predefined categories based on their content. It is the automated assignment of natural language texts to predefined categories. Text classification is the primary requirement of text retrieval systems, which retrieve texts in response to a user query, and text understanding systems, which transform text in some way such as producing summaries, answering questions or extracting data. Existing supervised learning algorithms to automatically classify text need sufficient documents to learn accurately. This paper presents a new algorithm for text classification using data mining that requires fewer documents for training. Instead of using words, word relation i.e. association rules from these words is used to derive feature set from pre-classified text documents. The concept of Naïve Bayes classifier is then used on derived features and finally only a single concept of Genetic Algorithm has been added for final classification. A system based on the proposed algorithm has been implemented and tested. The experimental results show that the proposed system works as a successful text classifier.

## 1. Introduction

There are numerous text documents available in electronic form. More and more are becoming available every day. Such documents represent a massive amount of information that is easily accessible. Seeking value in this huge collection requires organization; much of the work of organizing documents can be automated through data mining. The accuracy and our understanding of such systems greatly influence their usefulness [9]. The task of data mining is to automatically classify documents into predefined classes based on their content. Many algorithms have been developed to deal with automatic text classification [4]. With the existing algorithms, a number of newly established processes are involving in the automation of text classification [20]. The most common techniques used for this purpose include Association Rule Mining, Implementation of Naïve Bayes Classifier, Genetic Algorithm, Decision Tree and so on.

Association rule mining [1] finds interesting association or correlation relationships among a large set of data items [4]. The discovery of these relationships among huge amounts of transaction records can help in many decision making process. On the other hand, the Naïve Bayes classifier uses the maximum a posteriori estimation for learning a classifier. It assumes that the occurrence of each word in a document is conditionally independent of all other words in that document given its class [3]. Although the Naïve Bayes works well in many studies [7], it requires a large number of training documents for learning accurately. Genetic algorithm starts with an initial population which is created consisting of randomly generated rules. Each rule can be represented by a string of bits. Based on the notion of survival of the fittest, a new population is formed to consist of the fittest rules in the current

[1] Assistant Professor, Department of Computer Science and Engineering, Manarat International University, Dhaka-1212, Bangladesh
[2] Department of Computer Science and Engineering, International Islamic University Chittagong, Chittagong-4203, Bangladesh
[3] School of Communication, Independent University Bangladesh, Chittagong-4203, Bangladesh

population, as well as offspring of these rules. Typically, the fitness of a rule is assessed by its classification accuracy on a set of training examples.

This paper presents a new algorithm for text classification. Instead of using words, word relation i.e. association rules is used to derive feature set from pre-classified text documents. The concept of Naïve Bayes Classifier is then used on derived features and finally a concept of Genetic Algorithm has been added for final classification. A system based on the proposed algorithm has been implemented and tested. The experimental results show that the proposed system works as a successful text classifier.

The supervised learning algorithms still used to automatically classify text need sufficient documents to learn accurately while this proposed technique requires fewer documents for training. Here association rules from the significant words are used to derive feature set from pre-classified text documents. These rules are then used with the concept of Naïve Bayes classifier and during testing phase a concept of Genetic Algorithm has been added for final classification. Our observed experiment on this concept shows that the classifier build this way is more accurate than the existing text classification systems.

## 2. Previous Works

Researchers showed a number of dependable techniques for text classification. During our research we have tried to consider the most recently established efficient approaches that we have found as a result of our search. In this chapter we have tried to present a study on some research papers to give a view on three techniques to classify text.

### 2.1 Using Association Rule with Naïve Bayes Classifier

Research on Text Classification Using the Concept of Association Rule of Data Mining where Naive Bayes Classifier was used to classify text finally showed the dependability of the Naïve Bayes Classifier with Associated Rules [3][5]. But since this method ignores the negative calculation for any specific class determination in some cases accuracy may fall. As e.g. to classify a text it just calculates the probability of different class with the probability values of the matched set while ignoring the unmatched sets. As a result if test set matches with a rule set, which has weak probability to the actual class, may cause wrong classification.

Steps observed to classify text using association rule with Naïve Bayes classifier are [5]:

- ❑ Each abstracts used to train is considered as a transaction in the text data.
- ❑ The text data is cleaned by removing unnecessary words i.e. text data is filtered and subject related words are collected.
- ❑ Association rule mining is applied to the set of transaction data where each frequent.
- ❑ Word set from each abstract is considered as a single transaction.
- ❑ A large word set is generated with their occurrence frequency.
- ❑ Then Naïve Bayes classifier is used for probability calculation.
- ❑ Before classifying a new document the text data (abstract), target class of which is to be determined, is again preprocessed similar to the process applied to training data.
- ❑ Frequent words are viewed as word sets.
- ❑ Matching words set(s) or its subset (containing items more than one) in the list of word sets collected from training data with that of subset(s) (containing items more than one) of frequent word set of new document is searched.
- ❑ The corresponding probability values of matched word set(s) for each target class are collected.
- ❑ Last of all the probability values for each target class from Naïve Bayes classification theorem are calculated and the corresponding class of a new document is determined.

**2.2 Using Association Rule with Decision Tree**

According to the analyzed paper on Text classification using decision tree, an acceptable accuracy was obtained using 76% training data of total data set [16], while it is possible to achieve good accuracy using only 40 to 50% of total data sets as training data. Steps observed in this paper to implement the technique are [16]:

- ❑ Cleaning text.
- ❑ Implementation of Apriori algorithm to generate frequent candidate item set.
- ❑ Generation of association rules.
- ❑ Using a decision tree generator a decision tree is generated.
- ❑ Use of this tree to find out the associated class of new text document.

**2.3 With Genetic Algorithm**

Text Classification based on genetic algorithm showed satisfactory performance using 69% training data, but this process requires the time consuming steps to classify the texts [2][6]:

- ❑ Use of genetic algorithm to generate rules for text categorization. To avoid biasing effect by large number of documents or large sized document of a particular type, relative frequency of a word in all documents is used.
- ❑ Use of Roulette Wheel Selection method.
- ❑ Use of multipoint crossover. Both the number of crossover points and their locations as well as the chromosomes to participate in crossover is set random.
- ❑ Mutation probability of a bit (gene) is controlled depending on its corresponding word's weight.

## 3. Background Study

### 3.1 Data Mining

Data mining refers to extracting or "mining" knowledge from large amounts of data. It can also be named by "knowledge mining form data". Nevertheless, mining is a vivid term characterizing the process that finds a small set of precious nuggets from a great deal of raw material. There are many other terms carrying a similar or slightly different meaning to data mining, such as knowledge mining from databases, knowledge extraction, data/ pattern analysis, data archaeology, and data dredging.

Many people treat data mining as a synonym for another popularly used term, Knowledge Discovery in Databases, or KDD. Alternatively, data mining is also treated simply as an essential step in the process of knowledge discovery in databases.

The fast-growing, tremendous amount of data, collected and stored in large and numerous databases, has far exceeded our human ability for comprehension without powerful tools. In such situation we become data rich but information poor. Consequently, Important decisions are often made based not on the information-rich data stored in databases but rather on a decision maker's intuition, simply because the decision maker does not have the tools to extract the valuable knowledge embedded in the vast amounts of data. In addition, current expert system technologies rely on users or domain experts to manually input knowledge into knowledge bases. Unfortunately, this procedure is prone to biases and errors, and is extremely time consuming and costly. In such situation data mining tools can perform data analysis and may uncover important data patterns, contributing greatly to business strategies, knowledge bases, and scientific or medical research [7].

### 3.2 Association Rule

Association rule mining finds interesting association or correlation relationships among a large set of data items. In short association rule is based on associated relationships. The discovery of interesting association relationships among huge amounts of transaction records can help in many decision-making processes. Association rules are generated on the basis of two important terms namely minimum support threshold and minimum confidence threshold.

Let us consider the following assumptions to represent the association rule in terms of mathematical representation,

J = {i1, i2, . . . , im} be a set of items. Let D the task relevant data, be a set of database transactions where each transaction T is a set of items such that T $\subseteq$ J. Each transaction is associated with an identifier, called TID. Let A be a set of items. A transaction T is said to contain A if and only if A $\subseteq$ T. An association rule is an implication of the form A $\Longrightarrow$ B, where A $\subset$ J, B $\subset$ J, and A$\cap$B = $\Phi$. The rule A $\Longrightarrow$ B holds in the transaction set D with support s, where s is the percentage of transactions in D that contain A$\cup$B (i.e. both A and B). This is taken to be the probability, P (A$\cup$B). The rule A $\Longrightarrow$ B has confidence c in the transaction set d if c is the percentage of transaction in D containing A that also contain B. This is taken to be the conditional probability, P (B | A). That is,

$$\text{support } (A \Longrightarrow B) = P (A \cup B) \text{ and confidence } (A \Longrightarrow B) = P (B | A)$$

Association Rules that satisfy both a minimum support threshold and minimum confidence threshold are called strong association rules. A set of items is referred to as an itemset. In data mining research literature, "itemset" is more commonly used than "item set". An itemset that contains k items is a k-itemset. The occurrence frequency of an itemset is the number of transactions that contain the itemset. This is also known, simply as the frequency, support count, or count of the itemset. An itemset satisfies minimum support if the occurrence frequency of the itemset is greater than or equal to the product of minimum support and the total number of transactions in D. The number of transactions required for the itemset to satisfy minimum support is therefore referred to as the minimum support count. If an itemset satisfies minimum support, then it is a frequent itemset. The set of frequent k-itemsets is commonly denoted by Lk.

Association rule mining is a two-step process, which includes:

1. Find all Frequent Itemsets
2. Generate Strong Association Rules from the Frequent Itemsets

### 3.2.1 The Apriori Algorithm

Apriori is an influential algorithm for mining frequent itemsets for Boolean association rules. The name of the algorithm is based on the fact that the algorithm uses prior knowledge of frequent itemset properties.

Apriori employs an iterative approach known as a level-wise search, where k-itemsets are used to explore (k+1)-itemsets. First, the set of frequent 1-itemsets is found. This set is denoted by L1. L1 is used to find L2, the set of frequent 2-itemsets, which is used to find L3, and so on, until no more frequent k-itemsets can be found. The finding of each Lk requires one full scan of the database.

To understand how Apriori property is used in the algorithm, let us look at how Lk-1 is used to find Lk. A two step process is followed, consisting of join and prune actions:

### i) The Join Step:

To find Lk, a set of candidate k-itemsets is generated by joining Lk-1 with itself. This set of candidates is denoted by Ck. Let l1 and l2 be itemsets in Lk-1 then l1 and l2 are joinable if their first (k-2) items are in common, i.e., (l1[1] = l2[1]) . (l1[2] = l2 [2]) . . . . . . (l1[k-2]= l2[k-2]) . (l1[k-1]< l2 [k-1]).

### ii) The Prune Step:

Ck is the superset of Lk. A scan of the database to determine the count if each candidate in Ck would result in the determination of Lk (itemsets having a count no less than minimum support in Ck). But this scan and computation can be reduced by applying the Apriori property. Any (k-1)-itemsets that is not frequent cannot be a subset of a frequent k-itemset. Hence if any (k-1)-subset of a candidate k-itemset is not in Lk-1, then the candidate cannot be frequent either and so can be removed from Ck.

The algorithm is as follows

Input:          Database, D;
                        minimum support threshold, min_sup.

Output:        L, frequent itemsets in D.

Method:

(1)        L1=find_frequent_1-itemsets (D);
(2)        for (k=2; Lk-1 ≠ Φ; k++){
(3)            Ck = apriori_gen (Lk-1, min_sup);
(4)        for each transaction t    D { // scan D for counts
(5)            Ct = subset (Ck, t);
                                // get the subsets of t that are candidates
(6)            for each candidate c    Ct
(7)                    c.count++;
(8)        }
(9)            Lk = {c    Ck | c.count ≥ min_sup}
(10)}
(11)        return L = UkLk;

procedure apriori_gen
(Lk-1: frequent (k-1)-itemsets; min_sup: minimum support threshold)

(1)        for each itemset l1    Lk-1
(2)        for each itemset l2    lk-1
(3)            if (l1[1]=l2[1]). (l1[2]=l2 [2]) ... (l1[k-2]=l2[k-2]).(l1[k-1]<l2 [k-1]) then {
(4)            c = l1 | l2; // join step: generate candidates
(5)            if has_infrequent_subset (c,Lk-1) then
(6)                delete c; // prune step: remove unfruitful candidate
(7)            else add c to Ck;
(8)        }
(9)        return Ck;

procedure has_infrequent_subset
(c: candidate k-itemset; Lk-1: frequent (k-1)-itemsets); // use prior knowledge

(1)        for each (k-1)-subset s of c
(2)            if s ∉ Lk-1 then
(3)                return TRUE;
(4)        return FALSE;

### 3.2.1.1 Illustration of Apriori Algorithm

Let us consider an example of Apriori, based on the following transaction database, D of figure 4.1, with 9 transactions, to illustrate Apriori algorithm.

| TID | List of item_IDs |
|-----|------------------|
| T100 | l1, l2, l5 |
| T200 | l2, l4 |
| T300 | l2, l3 |
| T400 | l1, l2, l4 |
| T500 | l1, l3 |
| T600 | l2, l3 |
| T700 | l1, l3 |
| T800 | l1, l2, l3, l5 |
| T900 | l1, l2 , l3 |

**Figure 1.** Transactional Data

| Itemset | Sup. count |
|---------|-----------|
| {l1} | 6 |
| {l2} | 7 |
| {l3} | 6 |
| {l4} | 2 |
| {l5} | 2 |

**Figure 2.** Generation of C1

| Itemset | Sup. count |
|---------|-----------|
| {l1} | 6 |
| {l2} | 7 |
| {l3} | 6 |
| {l4} | 2 |
| {l5} | 2 |

**Figure 3.** Generation of L1

❑ In the first iteration of the algorithm, each item is a number of the set of candidate 1-itemsets, C1. The algorithm simply scans all of the transactions in order to count the number of occurrences of each item.

❑ Suppose that the minimum transaction support count required is 2 (i.e.; min_sup = 2/9 = 22%). The set of frequent 1-itemsets, L1, can then be determined. It consists of the candidate 1-itemsets satisfying minimum support.

❑ To discover the set of frequent 2-itemsets, L2, the algorithm uses L1 | L2 to generate a candidate set of 2-itemsets, C2.

❑ The transactions in D are scanned and the support count of each candidate itemset in C2 is accumulated.

| Itemset | Sup.count |
|---------|-----------|
| {l1, l2} | 4 |
| {l1, l3} | 4 |
| {l1, l4} | 1 |
| {l1, l5} | 2 |
| {l2, l3} | 4 |
| {l2, l4} | 2 |
| {l2, l5} | 2 |
| {l3, l4} | 0 |
| {l3, l5} | 1 |
| {l4, l5} | 0 |

**Figure 4.** Generation of C2

| Itemset | Sup.count |
|---------|-----------|
| {l1, l2} | 4 |
| {l1, l3} | 4 |
| {l1, l5} | 2 |
| {l2, l3} | 4 |
| {l2, l4} | 2 |
| {l2, l5} | 2 |

**Figure 5.** Generation of L2

| Itemset | Sup.count |
|---------|-----------|
| {l1, l2, l3} | 2 |
| {l1, l2, l5} | 2 |

**Figure 6.** Generation of C3

| Itemset | Sup.count |
|---------|-----------|
| {l1, l2, l3} | 2 |
| {l1, l2, l5} | 2 |

**Figure 7.** Generation of L3

❑ The set of frequent 2-itemsets, L2, is then determined, consisting of those candidate-itemsets in C2 having minimum support.

- ❑ The generation of the set of candidate 3-itemsets, C3 is observed in Fig 4.6 to Fig 4.7. Here C3 = L1 | L2 = {{l1, l2, l3}, {l1, l2, l5}, {l1, l3, l5}, {l2, l3, l5}, {l2, l4, l5}}. Based on the Apriori property that all subsets of a frequent itemset must also be frequent, we can determine that the four latter candidates cannot possibly be frequent.
- ❑ The transactions in D are scanned in order to determine L3, consisting of those candidate 3-itemsets in C3 having minimum support.
- ❑ The algorithm uses L3 | L3 to generate a candidate set of 4-itemsets, C4. Although the join results in {{l1, l2, l3, l5}}, this itemset is pruned since its subset {{l2, l3, l5}} is not frequent. Thus, C4 = {}, and the algorithm terminates.

### 3.3 Naive Bayes Classifier

Bayesian classification is based on Bayes theorem. A simple Bayesian classification namely the Naïve classifier is comparable in performance with decision tree and neural network classifiers. Bayesian classifiers have also exhibited high accuracy and speed when applied to large database.

Naïve Bayes classifier assumes that the effect of an attribute value on a given class is independent of the values of the other attributes. This assumption is called *class conditional independence*. It is made to simplify the computations involved and, in this sense, is considered "naïve" [7].

While applying Naïve Bayes classifier to classify text, each word position in a document is defined as an attribute and the value of that attribute to be the word found in that position. Here Naïve Bayes classification can be given by:

$$V_{NB} = \text{argmax } P(V_j) \prod P(a_j | V_j)$$

Here $V_{NB}$ is the classification that maximizes the probability of observing the words that were actually found in the example documents, subject to the usual Naïve Bayes independence assumption. The first term can be estimated based on the fraction of each class in the training data. The following equation is used for estimating the second term:

$$\frac{n_k + 1}{n + | \text{ vocabulary } |}$$

where n is the total number of word positions in all training examples whose target value is $V_j$, $n_k$ is the number of items that word is found among these n word positions, and | vocabulary | is the total number of distinct words found within the training data.

### 3.4 Genetic Algorithm

Genetic algorithms are a part of evolutionary computing, which is a rapidly growing area of artificial intelligence. As we can guess, genetic algorithms are inspired by Darwin's theory of evolution. Simply said, problems are solved by an evolutionary process resulting in a best (fittest) solution (survivor) - in other words, the solution is evolved [18].

According to Artificial Intelligence, Structures and Strategies for Complex Problem Solving, Fourth Edition, at page 471. Luger, George F. 2002. Harlow, England: Addison-Wesley:

*"Genetic algorithms are based on a biological metaphor: They view learning as a competition among a population of evolving candidate problem solutions. A 'fitness' function evaluates each solution to decide whether it will contribute to the next generation of solutions. Then, through operations analogous to gene transfer in sexual reproduction, the algorithm creates a new population of candidate solutions."*

In general, genetic algorithm starts with an initial population which is created consisting of randomly generated rules. Each rule can be represented by a string of bits. As a sample example, suppose the samples in a given training set are described by two boolean attributes, A1 and A2, and that there are two classes, C1 and C2. The rule "IF A1

AND NOT A2 THEN C2" can be encoded as the bit string "100", where the two leftmost bits represent attributes A1 and A2, respectively and the rightmost bit represents the class. Similarly, the rule "IF NOT A1 AND NOT A2 THEN C1" can be encoded as the bit string "001", If an attribute has k-values, where k>2,then k-bits may be used to encode the attribute's values. Classes can be encoded in a similar fashion.

Based on the notion of survival of the fittest, a new population is formed to consist of the fittest rules in the current population, as well as offspring of these rules. Typically, the fitness of a rule is assessed by its classification accuracy on a set of training examples.

Offspring are created by applying genetic operators such as crossover and mutation. In crossover, substrings from pairs of the rules are swapped to form new pairs of rules. In mutation, randomly selected bits in a rule's string are inverted. The process of generating new populations based on prior populations of rules continues until a population P "evolves" where each rule in P satisfies a pre-specified fitness threshold [7].

## 4. The Proposed Method

Our proposed method to classify text is an implementation of Association Rule with a combined use of Naïve Bayes Classifier and Genetic Algorithm. We have used the features of association rule to make association sets. On the other hand, to make a probability chart with prior probabilities we have used Naïve Bayes classifier's probability measurements. And in the last retrieval phase of test data we have implemented the positive-negative matching calculation observed in different researches [2][6].

One thing that has to be noticed is that Genetic Algorithm's conventional phases like crossover, mutation are not included in this method. The only thing of genetic algorithm that we used is the matching to the desired class and mismatching to the other classes. Here he associated word sets, which do not mach our considered class is treated as negative sets and others are positive.

### 4.1 The Proposed Algorithm

The following algorithm is applicable at the class determination phase of testing phase. That is after the probability table as well as association rules have been created for the training data, text preparation for the test data is done and then this algorithm is applied for the classification.

The algorithm is as follows:

*n = number of class,*
*m = number of associated sets*

1.   *For each class i = 1 to n do*
2.     *Set pval = 0, nval = 0, p = 0, n = 0*
3.   *For each set s = 1 to m do*
4.       *If the probability of the class (i) for the set (s) is maximum then*
            *increment pval else increment nval*
5.       *If 50% of the associated set s is matched with the keywords*
            *set do step 6 else do step7*
6.       *If maximum probability matches the class i then*
            *increment p*
7.       *If maximum probability does not match the class i*
            *increment n*
8.     *If (s<=m)*
            *go to step 3*
9.   *Calculate the percentage of matching in positive sets for the class i*
10.  *Calculate the percentage of not matching in negative sets for the class i*

11. *Calculate the total probability as the summation of the results obtained from step 9 and 10 and also the prior probability of the class i in set s*
12. *If (i<=n)*
    *go to step 1*
13. *Set the class having the maximum probability value as the result*

**4.2 Flow Chart of the Technique**

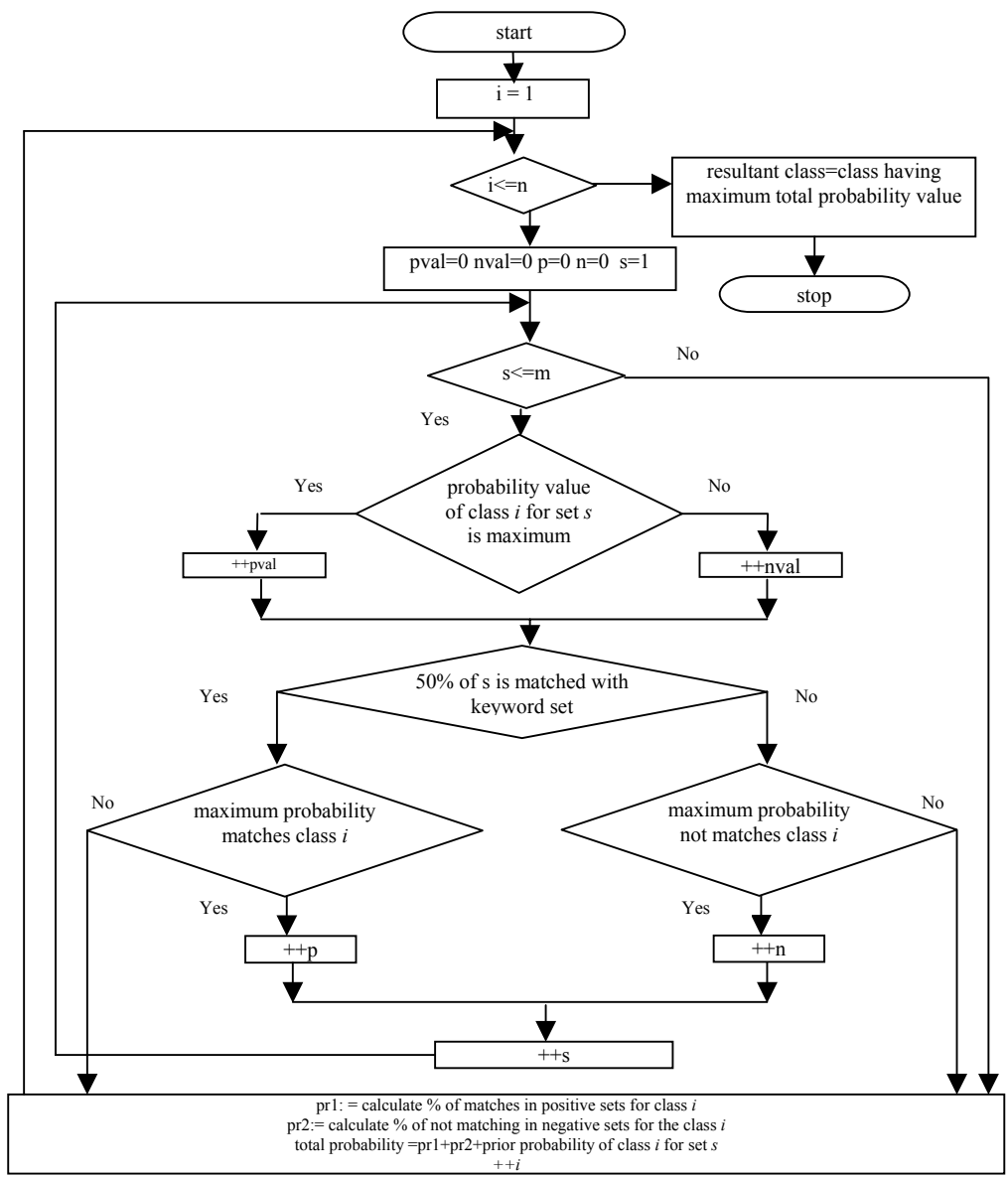

**Figure 8.** Flowchart of the Proposed Algorithm

# 5. Experimental Evaluation

## 5.1 Preparing Text for Classification

Abstracts from different research papers have been used to analyze the experiment. Five classes of papers from Physics(PH), Chemistry(CH), Algorithm (ALG), Educational Engineering (EDE) and Artificial Intelligence (AI) were considered for our experiment. We used a total of 296 abstracts (104 from Physics, 88 from Chemistry, 27 from Algorithm, 15 from Educational Engineering and 62 from AI).

**Table 1.** Distribution of Data to Make the Machine Intelligent

| Inserted Data to Train | | | | | | |
|---|---|---|---|---|---|---|
| Total% | Total Amount of Data | PH | CH | ALG | EDE | AI |
| 15 | 44 | 16 | 13 | 4 | 2 | 9 |
| 20 | 59 | 21 | 18 | 5 | 3 | 12 |
| 25 | 75 | 26 | 22 | 7 | 4 | 16 |
| 30 | 89 | 31 | 26 | 8 | 5 | 19 |
| 35 | 104 | 36 | 31 | 10 | 5 | 22 |
| 40 | 119 | 42 | 35 | 11 | 6 | 25 |
| 45 | 134 | 47 | 40 | 12 | 7 | 28 |
| 50 | 144 | 51 | 43 | 13 | 7 | 30 |
| 55 | 162 | 57 | 48 | 15 | 8 | 34 |

To make the raw text valuable, that is to prepare the text, we have considered only the keywords. That is unnecessary words and symbols are removed. For this keyword extraction process we dropped the common unnecessary words like am, is, are, to, from...etc. and also dropped all kinds of punctuations and stop words. Singular and plural form of a word is considered same. Finally, the remaining frequent words are considered as keywords.

Let an abstract:

*With respect to all algorithm perspective coding binary trees and an representation for well-formed parentheses strings. We present here the first Gray code and loopless generating algorithm for P-sequences, and extend them in a Gray code and a new loopless generating algorithm for well-formed parentheses strings.Given a connected graph G = (V,E) and a span-ning tree T of G, a fundamental cycle is a cycle resulting by adding an edge e ÍE - T to T. In this paper we estab-lish that the average length of fundamental cycles in a complete graph increases with the number of vertices. Also, given a simple cycle in a complete graph, the paper describes a method of calculating the number of spanning trees, with respect to which the cycle is a fundamental cycle.*

Keywords extracted from this abstract are:

*respect, algorithm, tree, well, formed, parenthese, string, gray, code, looples, generating, graph, fundamental, cycle, paper, complete, number*

This keyword extraction process is applied to all the abstracts.

## 5.2 Training Phase

### 5.2.1 Deriving Associated Word Sets

Each abstract is considered as a transaction in the text data. After pre-processing the text data association rule mining [1] is applied to the set of transaction data where each frequent word set from each abstract is considered as a single transaction. Using these transactions, we generated a list of maximum length sets applying the Apriori algorithm [1]. The support and confidence is set to 0.05 and 0.75 respectively. A list of the generated large word set for 55% of training data with their occurrence frequency is illustrated in Table 1.

**Table 2.** Word set with Occurrence Frequency

| Large Word Set Found | Number of Occurence in Documents | | | | |
|---|---|---|---|---|---|
| | **PH** | **CH** | **ALG** | **EDE** | **AI** |
| present, well, formed, parenthese, looples | | | 2 | | |
| spanning, tree | | | 2 | | |
| set, length | | | 2 | | |
| huffman, coding | | | 2 | | |
| generating, gray, code | | | 3 | | |
| educational, significant, study, education, level, student, learning | | | | 3 | |
| handicapped, more, different, environment, effect, working, motivation | | | | 2 | |
| difference, study | | | | 3 | |
| test, significant, difference | | | | 2 | |
| educational, significant, study, teacher | | | | 3 | |
| teacher, significant, study, student | | | | 3 | |
| education, level, significant, difference | | | | 3 | |
| school, teacher, education, level, significant | | | | 3 | |
| respect, job, level, significant, difference | | | | 2 | |
| three, coding, eZW, dimension, performance, lossy, compared, progressive | | | | | 2 |
| gain, dependencie | | | | | 2 |
| network, neural | | | | | 5 |
| channel, rate | | | | | 2 |
| rate, different | | | | | 3 |
| allocation, different | | | | | 2 |
| proposed, different | | | | | 2 |
| compression, progressive, method | | | | | 3 |
| waveleet, imagery, decoding, pixel | | | | | 3 |
| transform, imagery, decoding, pixel | | | | | 2 |
| communication, video, coder, constant, quality | | | | | 2 |
| lossles, image, compression, performance, lossy, progressive | | | | | 3 |
| even, number, spin, state | 2 | | | | |
| flow, black, hole, thermal | 3 | | | | |
| ground, dot | 2 | | | | |
| neutron, star | 3 | | | | |
| time, alpha | 3 | | | | |
| variation, line | 2 | | | | |
| analysi, high | 2 | | | | |
| main, sequence | 2 | | | | |
| quasar, luminosity | 3 | | | | |
| above, luminosity | 2 | | | | |
| consistent, propertie | 2 | | | | |
| binary, mas | 2 | | | | |
| microlensing, caustic | 2 | | | | |
| velocity, seen | 2 | | | | |
| cold, dark | 2 | | | | |
| obtained, alpha, line | 2 | | | | |
| nuclear, collapse, simulation | 2 | | | | |
| giant, planet, due | 2 | | | | |
| calculate, raman, response, fifth, order | | 2 | | | |
| scalar, included, approximation, dielectric, function | | 2 | | | |
| dirac, fock | | 2 | | | |
| hartree, fock | | 3 | | | |
| excited, using | | 3 | | | |
| site, ion | | 2 | | | |
| geometry, simulation | | 3 | | | |
| hyperpolarizability, second | | 4 | | | |
| carbon, nanotube | | 3 | | | |
| strong, laser | | 2 | | | |
| many, body | | 2 | | | |
| third, fifth, order | | 3 | | | |
| third, order, calculated | | 3 | | | |
| hydrogen, model, agreement | | 2 | | | |
| quadrupole, moment, accurate | | 2 | | | |
| triplet, mode, geometry | | 2 | | | |
| polarization, solvent, charge | | 2 | | | |
| local, exchange, current | | 3 | | | |
| frequency, classical, water | | 2 | | | |
| raman, response, third, order | | 3 | | | |
| raman, response, order, effect | | 3 | | | |
| electronic, state, interest, transition | | 2 | | | |
| kohn, vignale, within, local | | 3 | | | |
| describe, exchange, current, vignale | | 2 | | | |
| rev, lett, van, improvement | | 2 | | | |

**5.2.2 Setting Associated Word Set with Probability Value**

To use the Naive Bayes classifier for probability calculation the generated associated sets are required. The calculation of first term of this classifier is based on the fraction of each target class in the training data.

From the generated word set after applying association mining on training data we have found the following information:

total number of word set = 69
    total number of word set from Physics (PH) = 17
    total number of word set from Chemistry (CH) = 25
    total number of word set from Algorithm (ALG) = 5
    total number of word set from Educational Engineering (EDE) = 9
    total number of word set from Artificial Intelligence (AI) = 12

Prior probability we had for PH, CH, ALG, EDE and AI are 0.26, 0.36, 0.07, 0.13 and 0.17 respectively. Then the second term is calculated according to the equation (1). The probability values of word set are listed in Table 2.

**Table 3.** Word set with Probability Value

| Large Word Set Found | Pobability | | | | |
|---|---|---|---|---|---|
| | **PH** | **CH** | **ALG** | **EDE** | **AI** |
| present, well, formed, parenthese, looples | 0.013514 | 0.013514 | 0.040541 | 0.013514 | 0.013514 |
| spanning, tree | 0.013514 | 0.013514 | 0.040541 | 0.013514 | 0.013514 |
| set, length | 0.013514 | 0.013514 | 0.040541 | 0.013514 | 0.013514 |
| huffman, coding | 0.013514 | 0.013514 | 0.040541 | 0.013514 | 0.013514 |
| generating, gray, code | 0.013514 | 0.013514 | 0.054054 | 0.013514 | 0.013514 |
| educational, significant, study, education, level, student, learning | 0.012821 | 0.012821 | 0.012821 | 0.051282 | 0.012821 |
| handicapped, more, different, environment, effect, working, motivation | 0.012821 | 0.012821 | 0.012821 | 0.038462 | 0.012821 |
| difference, study | 0.012821 | 0.012821 | 0.012821 | 0.051282 | 0.012821 |
| test, significant, difference | 0.012821 | 0.012821 | 0.012821 | 0.038462 | 0.012821 |
| educational, significant, study, teacher | 0.012821 | 0.012821 | 0.012821 | 0.051282 | 0.012821 |
| teacher, significant, study, student | 0.012821 | 0.012821 | 0.012821 | 0.051282 | 0.012821 |
| education, level, significant, difference | 0.012821 | 0.012821 | 0.012821 | 0.051282 | 0.012821 |
| school, teacher, education, level, significant | 0.012821 | 0.012821 | 0.012821 | 0.051282 | 0.012821 |
| respect, job, level, significant, difference | 0.012821 | 0.012821 | 0.012821 | 0.038462 | 0.012821 |
| three, coding, eZW, dimension, performance, lossy, compared, progressive | 0.012346 | 0.012346 | 0.012346 | 0.012346 | 0.037037 |
| gain, dependencie | 0.012346 | 0.012346 | 0.012346 | 0.012346 | 0.037037 |
| network, neural | 0.012346 | 0.012346 | 0.012346 | 0.012346 | 0.074074 |
| channel, rate | 0.012346 | 0.012346 | 0.012346 | 0.012346 | 0.037037 |
| rate, different | 0.012346 | 0.012346 | 0.012346 | 0.012346 | 0.049383 |
| allocation, different | 0.012346 | 0.012346 | 0.012346 | 0.012346 | 0.037037 |
| proposed, different | 0.012346 | 0.012346 | 0.012346 | 0.012346 | 0.037037 |
| compression, progressive, method | 0.012346 | 0.012346 | 0.012346 | 0.012346 | 0.049383 |
| waveleet, imagery, decoding, pixel | 0.012346 | 0.012346 | 0.012346 | 0.012346 | 0.049383 |
| transform, imagery, decoding, pixel | 0.012346 | 0.012346 | 0.012346 | 0.012346 | 0.037037 |
| communication, video, coder, constant, quality | 0.012346 | 0.012346 | 0.012346 | 0.012346 | 0.037037 |
| lossles, image, compression, performance, lossy, progressive | 0.012346 | 0.012346 | 0.012346 | 0.012346 | 0.049383 |
| even, number, spin, state | 0.034483 | 0.011494 | 0.011494 | 0.011494 | 0.011494 |
| flow, black, hole, thermal | 0.045977 | 0.011494 | 0.011494 | 0.011494 | 0.011494 |
| ground, dot | 0.034483 | 0.011494 | 0.011494 | 0.011494 | 0.011494 |
| neutron, star | 0.045977 | 0.011494 | 0.011494 | 0.011494 | 0.011494 |
| time, alpha | 0.045977 | 0.011494 | 0.011494 | 0.011494 | 0.011494 |
| variation, line | 0.034483 | 0.011494 | 0.011494 | 0.011494 | 0.011494 |
| analysi, high | 0.034483 | 0.011494 | 0.011494 | 0.011494 | 0.011494 |
| main, sequence | 0.034483 | 0.011494 | 0.011494 | 0.011494 | 0.011494 |
| quasar, luminosity | 0.045977 | 0.011494 | 0.011494 | 0.011494 | 0.011494 |
| above, luminosity | 0.034483 | 0.011494 | 0.011494 | 0.011494 | 0.011494 |
| consistent, propertie | 0.034483 | 0.011494 | 0.011494 | 0.011494 | 0.011494 |
| binary, mas | 0.034483 | 0.011494 | 0.011494 | 0.011494 | 0.011494 |
| microlensing, caustic | 0.034483 | 0.011494 | 0.011494 | 0.011494 | 0.011494 |
| velocity, seen | 0.034483 | 0.011494 | 0.011494 | 0.011494 | 0.011494 |
| cold, dark | 0.034483 | 0.011494 | 0.011494 | 0.011494 | 0.011494 |
| obtained, alpha, line | 0.034483 | 0.011494 | 0.011494 | 0.011494 | 0.011494 |
| nuclear, collapse, simulation | 0.034483 | 0.011494 | 0.011494 | 0.011494 | 0.011494 |
| giant, planet, due | 0.034483 | 0.011494 | 0.011494 | 0.011494 | 0.011494 |
| calculate, raman, response, fifth, order | 0.010638 | 0.031915 | 0.010638 | 0.010638 | 0.010638 |
| scalar, included, approximation, dielectric, function | 0.010638 | 0.031915 | 0.010638 | 0.010638 | 0.010638 |
| dirac, fock | 0.010638 | 0.031915 | 0.010638 | 0.010638 | 0.010638 |
| hartree, fock | 0.010638 | 0.031915 | 0.010638 | 0.010638 | 0.010638 |
| excited, using | 0.010638 | 0.042553 | 0.010638 | 0.010638 | 0.010638 |
| site, ion | 0.010638 | 0.031915 | 0.010638 | 0.010638 | 0.010638 |
| geometry, simulation | 0.010638 | 0.042553 | 0.010638 | 0.010638 | 0.010638 |
| hyperpolarizability, second | 0.010638 | 0.053191 | 0.010638 | 0.010638 | 0.010638 |
| carbon, nanotube | 0.010638 | 0.042553 | 0.010638 | 0.010638 | 0.010638 |
| strong, laser | 0.010638 | 0.031915 | 0.010638 | 0.010638 | 0.010638 |
| many, body | 0.010638 | 0.031915 | 0.010638 | 0.010638 | 0.010638 |
| third, fifth, order | 0.010638 | 0.042553 | 0.010638 | 0.010638 | 0.010638 |
| third, order, calculated | 0.010638 | 0.042553 | 0.010638 | 0.010638 | 0.010638 |
| hydrogen, model, agreement | 0.010638 | 0.031915 | 0.010638 | 0.010638 | 0.010638 |
| quadrupole, moment, accurate | 0.010638 | 0.031915 | 0.010638 | 0.010638 | 0.010638 |
| triplet, mode, geometry | 0.010638 | 0.031915 | 0.010638 | 0.010638 | 0.010638 |
| polarization, solvent, charge | 0.010638 | 0.031915 | 0.010638 | 0.010638 | 0.010638 |
| local, exchange, current | 0.010638 | 0.042553 | 0.010638 | 0.010638 | 0.010638 |
| frequency, classical, water | 0.010638 | 0.031915 | 0.010638 | 0.010638 | 0.010638 |
| raman, response, third, order | 0.010638 | 0.042553 | 0.010638 | 0.010638 | 0.010638 |
| raman, response, order, effect | 0.010638 | 0.042553 | 0.010638 | 0.010638 | 0.010638 |
| electronic, state, interest, transition | 0.010638 | 0.031915 | 0.010638 | 0.010638 | 0.010638 |
| kohn, vignale, within, local | 0.010638 | 0.042553 | 0.010638 | 0.010638 | 0.010638 |
| describe, exchange, current, vignale | 0.010638 | 0.031915 | 0.010638 | 0.010638 | 0.010638 |
| rev, lett, van, improvement | 0.010638 | 0.031915 | 0.010638 | 0.010638 | 0.010638 |

## 5.3 Testing Phase

### 5.3.1 Classifying a New Document

Let an abstract as a text to test:

*The dielectric function of heavy nonmetallic crystals are studied within a relativistic framework using the ADF-BAND program package. The calculations are based on the work that has been done to calculate the dielectric response of nonmetallic crystals in article [7]. The starting point of the relativistic corrections is the Dirac equation in an quasi-static electric field. As the Dirac equation is a four-component equation it is first reduced to a two-component equation with the Foldy-Wouthuysen transformation. The then obtained two-component Dirac-Hamiltonian is then used to find (after some treatments of this Hamiltonian) an expression for the matrixelements required With these matrixelements the dielectric function can be evaluated, but now relativistically corrected. The obtained relativistic corrected dielectric function was finally evaluated for some light crystals; C,Si,GaAs and He and for heavier crystals asto see if the relativistic corrections indeed improve on the dielectric function of the studied crystals in article [7]. The heavy crystals with large errors as compared to experiment in article [7] were studied. The expectation is that for elements with an atomic number greater or equal to 50 ( Z ¥ 50) the relativistic corrections become important.*

The extracted sets of keywords of a new abstract are:

*dielectric, function, heavy, nonmetallic, crystal, studied, relativistic, article, correction, dirac, equation, component, two, obtained, hamiltonian, some, matrixelement, evaluated, corrected*

All these keywords are sent to the last classifier machine where for each class a common circle runs. As a result of this run a probability is obtained for each class. As for example for the class Chemistry the given set gives:

pval=25, nval=44, p=2, n=43.
Now the probability of CH = ((p*100)/pval) + ((n*100)/nval) + prior probability of CH
$\qquad$ = ((2*100)/25) + ((43*100)/44) + 0.36
$\qquad$ = 106.09

For this set of keywords,
$\qquad$ Calculated Probability for class PH=101.89 $\qquad$ Calculated Probability for class CH=106.09
$\qquad$ Calculated Probability for class ALG=95.38 $\qquad$ Calculated Probability for class EDE=95.13
$\qquad$ Calculated Probability for class AI=94.91

From the above result we found the abstract belongs to class Chemistry (CH).

**Table 4**. Accuracy Regarding Different Set of Test Data

| Inserted Data To Test As Input | | | | | | Accuracy | | | | | | |
|---|---|---|---|---|---|---|---|---|---|---|---|---|
| Total% | Total Amount of Data | PH | CH | ALG | EDE | AI | Accurate Amount of Data at Different Classes | | | | | Total Amount of Data Found Accurate | % of Accuracy |
| | | | | | | | PH | CH | ALG | EDE | AI | | |
| 45 | 252 | 88 | 40 | 12 | 7 | 28 | 31 | 27 | 12 | 7 | 13 | 90 | 79.61 |
| 50 | 237 | 83 | 45 | 14 | 8 | 32 | 41 | 35 | 14 | 8 | 22 | 120 | 67.67 |
| 55 | 221 | 78 | 48 | 15 | 8 | 34 | 30 | 35 | 15 | 8 | 22 | 110 | 67.90 |
| 60 | 207 | 73 | 53 | 16 | 9 | 37 | 32 | 39 | 15 | 9 | 25 | 120 | 67.80 |
| 65 | 192 | 68 | 57 | 17 | 10 | 40 | 37 | 49 | 14 | 10 | 24 | 134 | 69.79 |
| 70 | 177 | 62 | 62 | 19 | 10 | 43 | 23 | 43 | 18 | 9 | 25 | 118 | 57.00 |
| 75 | 162 | 57 | 66 | 20 | 11 | 46 | 19 | 47 | 18 | 10 | 25 | 119 | 53.85 |
| 80 | 152 | 53 | 70 | 22 | 12 | 50 | 4 | 60 | 0 | 0 | 10 | 74 | 31.78 |
| 85 | 134 | 47 | 75 | 23 | 13 | 53 | 20 | 65 | 0 | 0 | 0 | 85 | 33.73 |

**5.4  Graphical Representation of the Experiment**

**Table 5.** Percentage of Accuracy Vs Percentage of Training Data

| % of Training Data | % of Accuracy |
|---|---|
| 10 | 31 |
| 15 | 34 |
| 20 | 32 |
| 25 | 54 |
| 30 | 57 |
| 35 | 70 |
| 40 | 68 |
| 45 | 68 |
| 50 | 80 |
| 55 | 68 |

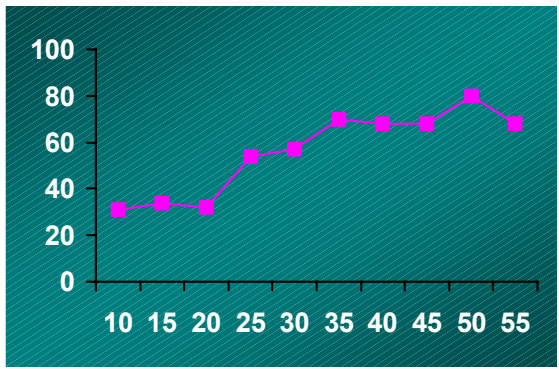

**Figure 9.** Accuracy Vs Training Curve of the Experimental Data

At the beginning of the experiment we started with 10% of the total data to train, which showed unsatisfactory accuracy. Then we increased training data to 20%, which showed development in accuracy. Next as we increased the percentage of training data accuracy became more desirable. We checked up to 55% training data. In this process, considering overall accuracy we choose 79% accuracy i.e., 50% training data as the best.

## 5.5 Comparative Study

In this section we have tried to represent comparative presentations in different point of views. We studied three thesis papers for the comparison purpose.

### 5.5.1  Association Rule and Naïve Bayes Classifier

The following results are found using the same data sets for both Association Rule with Naive Bayes Classifier and proposed method. The result shows that proposed approach work well using only 50% Training data.

**Table 6.** Comparison of Proposed Method with Text Classifier
using Association Rule and Naïve Bayes Classifier

| % of Training Data | % of Accuracy | |
|---|---|---|
| | **Association Rule with Naïve Byes Classifier** | **Proposed Method** |
| 10 | 40 | 30 |
| 20 | 17 | 32 |
| 30 | 42 | 57 |
| 40 | 60 | 68 |
| 50 | 32 | 80 |

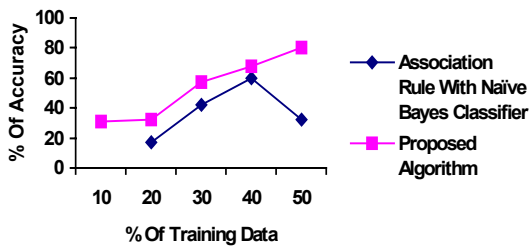

**Figure 9.** Accuracy Vs Training Curve Representing Comparative Output

### 5.5.2  Association Rule Based Decision Tree

In text categorization using association rule based decision tree [16] 76 % data set of the total 33 data set was used to train and 13% error observed. On the other hand using only 50% data as training the proposed algorithm can able to classify text with 80% accuracy rate. The major problem of decision tree based classified [10] [16] is that, this system is totally failed to categorize a class. Our proposed technique shows better performance even with 8 times larger data sets.

### 5.5.3  Genetic Algorithm

Researchers showed 68% accuracy using the concept of genetic algorithm with 31% test data [6] while our technique is better both in accuracy and % of test data. Moreover it required processing for each class during training. But our proposed algorithm does not require such process during training phase and hence reduces time.

**Table 7.** Comparison of Proposed Method with Text Classifier

using Decision Tree and Genetic Algorithm

| Technique | (%) Training Data | (%) Accuracy |
|---|---|---|
| Association Rule Based Decision Tree | 76 | 87 |
| Genetic Algorithm | 69 | 68 |
| Proposed Algorithm | 50 | 80 |

### 5.5.4 Overall Comparative Presentation

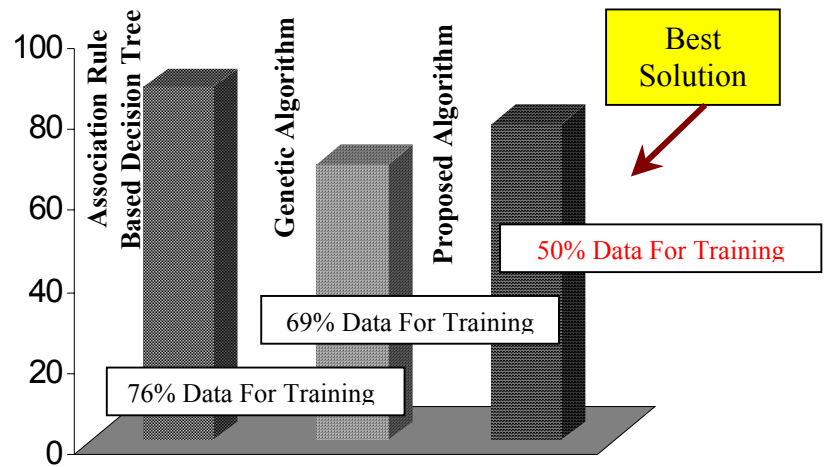

**Figure 10.**   Percentage of Accuracy Found for Different Researches

## 6. Limitations and Future Work

As we have observed in this method better accuracy is found with increasing confidence up to 0.75. The machine will be more effective if the training set is set in such a way that it generates more sets. That is training set with all the different sections of total data can give more dependable result. Moreover, if Frequent Pattern (FP) Growth tree could be formed time would be shorter enough [7]. Though the experimental results are quite encouraging, it would be better if we work with larger data sets with more classes. Addition of more features from Genetic Algorithm can make our technique to classify text more developed. In near future, we will try to apply this algorithm for emerging pattern identification.

## 7. Conclusion

This paper presented an efficient technique for text classification. The existing techniques require more data for training as well as the computational time of these techniques is also large. In contrast to the existing algorithms, the proposed hybrid algorithm requires less training data and less computational time. In spite of the randomly chosen training set we achieved 90% accuracy for 50% training data. Though the experimental results are quite encouraging, it would be better if we work with larger data sets with more classes.